\documentclass[aps,pre,twocolumn]{revtex4} 

\usepackage{epsfig}

\begin{document}

\title{Cluster phases of membrane proteins}

\author{Nicolas Destainville}
\affiliation{Laboratoire de Physique Th\'eorique~-- IRSAMC, UMR CNRS-UPS 5152,
Universit\'e de Toulouse, 31062 Toulouse Cedex 9, France.}

\date{\today}

\begin{abstract}
A physical scenario accounting for the existence of size-limited
submicrometric domains in cell membranes is proposed. It is based on
the numerical investigation of the counterpart, in lipidic membranes where
proteins are diffusing, of the recently discovered cluster phases in
colloidal suspensions. I demonstrate that the interactions between
proteins, namely short-range attraction and longer-range repulsion,
make possible the existence of stable small clusters. The consequences
are explored in terms of membrane organization and diffusion
properties.  The connection with lipid rafts is discussed
and the apparent protein diffusion coefficient as a function of
their concentration is analyzed.
\end{abstract}

\maketitle

\section{Introduction}

The question of membrane functional organization is a key issue in
modern cell biology \cite{1,2,3,4,5,6,6b,Goncalves06,7}. The central
problematics is to establish the relationship between (dynamical)
organization and functions of the different constituents of the
membranes. In this context, many microscopy techniques are implemented
to have access, with the highest possible spatial and temporal
resolutions, to the distribution and dynamics of membrane
proteins and lipids, in particular to their diffusive properties in
connection with their crowded environnement.

Using these techniques, there exist a large variety of situations, in
live cells \cite{1,3,6b,7,8,9,sieber06} or in model membranes \cite{4,10},
where membrane proteins are found in oligomers or small clusters. This
co-localization is supposed to facilitate encounters of partners in
different processes, such as complexes involved in signal transduction
\cite{1,2,3}. Lipid rafts~\cite{2,Edidin03,Fielding06} are usually
invoked to account for co-localization. These membrane submicrometric
domains enriched in certain lipids ({\em e.g.} cholesterol and
sphingolipids) are supposed to recruit proteins having a higher
affinity for their composition. They are believed to ensue from a
lipidic micro-phase separation~\cite{2,Edidin03,Fielding06}. However,
a consensus has not yet been reached to explain why the separation
process stops and domains remain size-limited~\cite{Fielding06}.  And
understanding the relationship between these structural patterns and
diffusion of their constituents remains topical~\cite{1,2,5,6,9}.

At the same time, there is an increasing interest in colloid science
for systems presenting a cluster phase. It is the fruit of a
competition between a short-range attraction ({\em e.g.} a depletion
interaction) which favors clusterization, and a longer-range repulsion
({\em e.g.} electrostatic) which prevents a complete phase separation,
because when clusters grow, their repulsion also grows and the
repulsion barrier cannot be passed by thermal activation anymore. The
result is an equilibrium phase with small clusters of
concentration-dependent size. It has been suggested \cite{11} that the
existence of cluster phases is not a singular behavior in a specific
system. They indeed occur in different physical systems ({\em e.g.}
colloids, star polymers, proteins) and for a large variety of
interactions \cite{11,Segre01,12,13,14,15,Sear99,17,16}.  Such
patterns have already been experimentally observed or simulated in
2D~\cite{10,Sear99,16,Seul95}.

I propose, as an alternative paradigm, to re-interpret the aggregation
of membrane proteins in terms of two-dimensional cluster phases,
because membrane proteins, like colloids, interact with energies of
the order of magnitude of the physiological thermal energy $k_BT$ ($T
\approx 310$~K)~\cite{2,19,20,Nielsen98,Gil98}. I discuss in this
framework the mechanism driving the formation of so-called rafts. I
propose that proteins spontaneously congregating in small clusters
instead of constituting a ``gas'' of independently diffusing
inclusions, they promote the formation of nano-domains in
membranes. The sub-micrometric, limited size of these nano-domains now
appears naturally in this scenario, as a result of the competition
between attraction and repulsion. By contrast to the lipid raft
scenario where domains result from a lipidic micro-phase separation
and then recruit specific proteins, the present mechanism proposes
that domain formation is mainly driven by protein interactions. Recent
experiments in live cells support this point of
view that protein-protein interactions are necessary to induce
clustering, whereas lipids alone are not sufficient in the system
studied~\cite{sieber06}.  I also relate the protein diffusive
properties as a function of their concentration to the limited size of
clusters, itself depending on concentration. I demonstrate that the
mean long-term diffusion coefficient $D$ of proteins decreases when
the mean cluster size grows: $D \propto 1/\langle n \rangle$ where
$\langle n \rangle$ is the mean cluster size (its number of
proteins). I also anticipate that $\langle n \rangle$ grows with the
protein concentration $\phi$ \cite{10}. Thus $D$ also depends on
$\phi$~\cite{9}. From this point of view, contact is made with prior
experiments, where $D$ was found to decrease like $1/\phi$~\cite{18}
or that were interpreted by appealing to such a behavior \cite{5}. I
propose a simple scenario leading to this law $D \sim 1/\phi$,
requiring that $\langle n \rangle \propto \phi$. Previous attempts to
address this issue, based on low $\phi$ expansions of $D$, are not
adapted to catch the physical mechanisms capable of accounting for
this behavior~\cite{Dean04}.

To give substance to and to support this cluster-phase scenario, I
first discuss available experimental data. In Ref.~\cite{10},
proteoliposomes of egg lecithin and bacteriorhodopsin (BR) are
observed by freeze-fracture electron microscopy. BR is found to
aggregate in small clusters (called ``particles''), the mean size of
which, $\langle n \rangle$, depends on BR concentration $\phi$. In
Table I (column 5), it is found that $\langle n \rangle \propto \phi$
for $1/\phi \in [90,300]$ (in mol./mol.  lipid to protein ratio), in
complete agreement with my arguments. In addition, the Fig.~4 of this
reference shows unambiguously that this behavior comes from the bimodal
character of the cluster size distribution: the balance between
monomers ($n=1$) and multimers ($n>1$) make possible the cluster
density to be nearly constant while the density of protein grows. I
will confirm below that it is exactly what is observed in
simulations. Moreover, the diffusion constant $D$ of BR is precisely
found to decay like $1/\phi$ in the same interval in Ref.~\cite{18},
as anticipated above. At the end of this article, I shall propose
experiments to validate definitely this scenario.

\section{SHORT AND LONGER-RANGE INTERACTIONS}

There exist several short-range attractive forces between proteins embedded in
membranes, each with a range of a few nanometers and a binding
energy of order $k_BT$. They first of all consist of a depletion interaction
due to the 2D osmotic pressure of lipids on proteins,
which tends to bring them closer when they are about a nanometer apart
\cite{19}. There also exist hydrophobic mismatch interactions between
proteins, the hydrophobic core of which does not match the width of the
membrane~\cite{2}. The energy cost of the subsequent membrane
deformation increases with the distance between two identical
proteins, thus resulting in an attractive force. The energy scale is
of order $k_BT$~\cite{Nielsen98}. In membranes with several lipid
species, another scenario leads to attractive forces: proteins recruit
in their neighborhood lipids which match best their hydrophobic
core. The closer the proteins, the more energetically favorable the
configuration. Binding energies are also of order $k_BT$ or
larger~\cite{2,Gil98}. A protein-driven mechanism for domain formation
invoking such forces has been proposed~\cite{2,Gil98}, but the limited
domain size due to additional repulsive forces has not been
discussed in this context.

Membrane inclusions are also affected by longer-range repulsive
forces. Electrostatic repulsion between like charged proteins is
usually considered as negligible because it is screened beyond a few
nanometers in physiological conditions: at physiological ionic
strength $I_\varphi\sim 0.1$~M, the Debye screening length is of the
order of 1~nm~\cite{Hunter}. Only proteins with (unreasonable) charges
of several hundreds of elementary charges can give a repulsion of a
fraction of $k_BT$ at 10~nm. By contrast, there exist specific repulsions
due to the deformation that proteins impose on the membrane when they
are not strictly speaking cylindrical inclusions but conical ones or
peripheral proteins \cite{20}. For example, using the formulae of this
reference for transmembrane proteins with a moderate contact angle of
$10^{\circ}$, one finds that the repulsive energy barrier at 10~nm is
0.10~$k_BT$ for a typical bending rigidity $\kappa=100 k_BT$. 
For example, rhodopsin has a contact angle larger than 
$10^{\circ}$~\cite{Deisen}.

Thus the ingredients for the existence of cluster phases are present
in assemblies of membrane proteins and cluster phases should
generically exist in cell membranes. Below, I shall take a typical
binding energy of $-4$~$k_BT$~\cite{Segre01} and an energy barrier of
0.1~$k_BT$ at about 10~nm.

\section{CLUSTER DIFFUSION}

\begin{figure}
\begin{center}
\parbox{4cm}{\psfig{figure=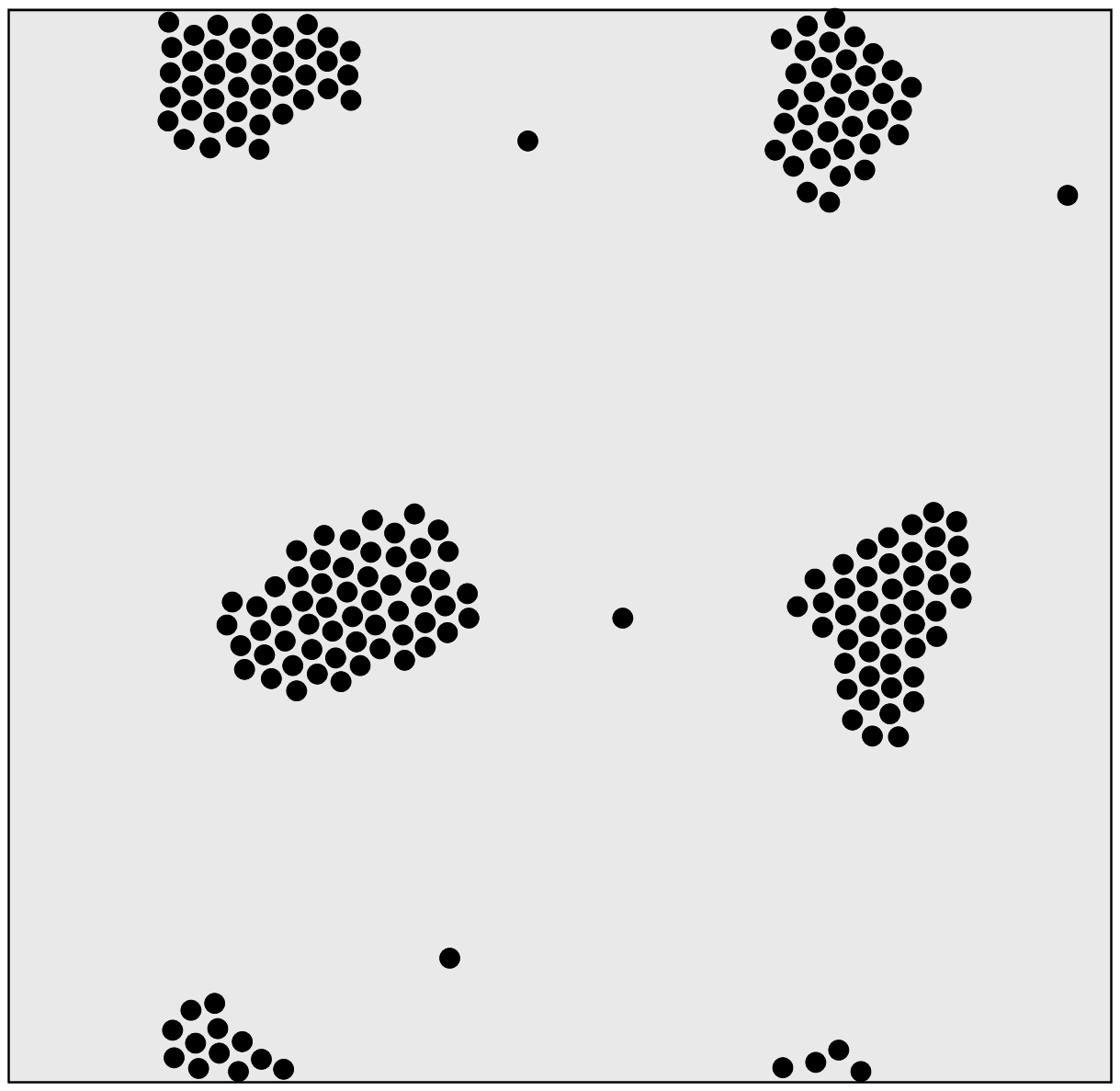,height=4.2cm,width=4.2cm} 
\hfill $t=31.0$~ms}~\parbox{4cm}{\psfig{figure=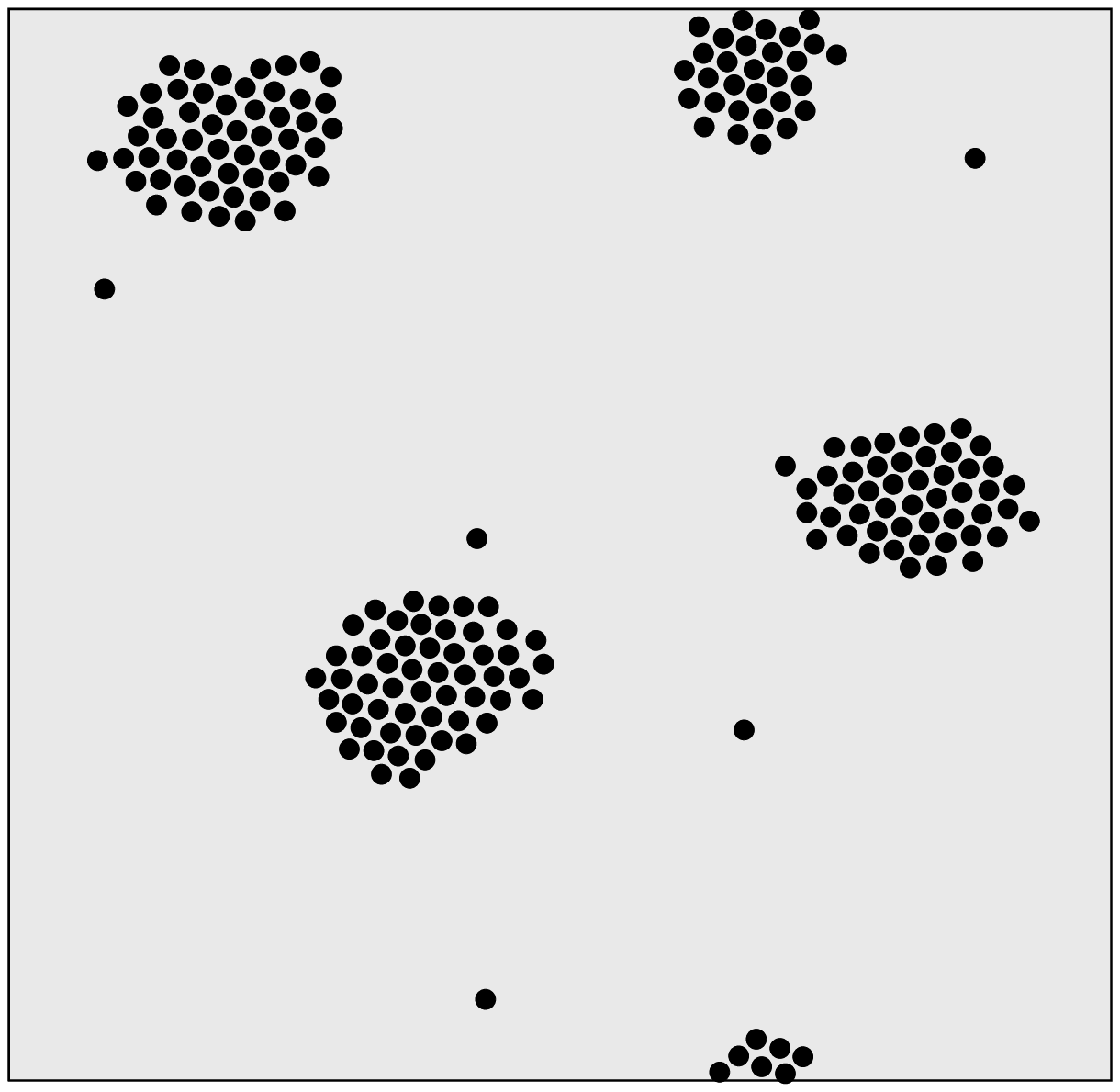,height=4.2cm,width=4.2cm}
\hfill $t=31.5$~ms}
\end{center}
\caption{Two snapshots of the cluster phase at $\phi = 0.1$ ($N = 200$
proteins, box side $a = 0.25$~$\mu$m, periodic boundary conditions);
The time delay between both snapshots is 0.5~ms. Clusters diffuse
slowly and appear non-rigid, deformable and fluctuating.
\label{figure}}
\end{figure}

First I study the diffusive properties of particles in cluster
phases. I consider an isolated cluster of $n$ proteins, modeled as
an assembly of interacting Langevin particles with ``bare'' diffusion
coefficient $D_0$ (the diffusion coefficient at vanishing
concentration). The center of mass of the assembly diffuses with a
coefficient $D_0/n$, because the clusters considered are not rigid
entities but loosely bound, fluctuating ones in which the proteins
diffuse \cite{5} (this property will be confirmed below). If clusters
interact weakly because they are sufficiently far away
(Fig.~\ref{figure}), the long-term diffusion coefficient of each
protein of the cluster is also equal to $D_0/n$ \cite{5}. If clusters
contain $\langle n \rangle$ proteins on average (counting a monomer as
a cluster with $n$ = 1), then the mean long-term diffusion coefficient
$D=D_0/\langle n \rangle$: if the system contains $N$ proteins
\begin{equation}
D \equiv \frac{1}{N} \sum_{c=1}^{N_c} n(c) \; \frac{D_0}{n(c)}, 
\end{equation}
where $N_c=N/\langle n\rangle$ denotes the number of clusters, because
a cluster $c$ contains $n(c)$ proteins that diffuse each with a
diffusion coefficient $D_0/n(c)$. Thus $D = D_0 N_c / N = D_0/\langle
n \rangle$. If $R$ is the average cluster radius, then $D \propto
1/R^2$. Such an experimental behavior of $D$ with cluster size has
already been observed~\cite{4}, but has been left unexplained.

Now, in 3D, it has been shown analytically \cite{12,15,17}, and
measured experimentally \cite{11,Segre01,13}, that $\langle n \rangle$
is proportional to the particle concentration $\phi$. My purpose here
is not to demonstrate such a relation, but simply to remark its
validity in a wide range of situations in 3D and to anticipate its
equivalent in 2D. In addition to the numerical evidence presented
below, further calculations, appealing for example to the theory of
micellization~\cite{Foret}, will be necessary to confirm this last
point. They go beyond the scope of the present numerical paper. If
$\langle n \rangle \propto \phi$, then the effective diffusion
coefficient of proteins in a cluster phase is inversely proportional
to their concentration:

\begin{equation} 
D = Const./\phi. 
\end{equation}

\section{MONTE CARLO SIMULATIONS}

{\em A priori}, these mechanisms should be valid only at low $\phi$,
where separate clusters interact weakly and diffuse independently. To
test further their relevance in 2D and at higher $\phi$, I
have performed Monte Carlo simulations of systems of $N=100$ particles
($N=200$ for the highest density considered $\phi=0.1$), interacting {\em via}
pairwise potentials displaying a hard-core repulsion, a short-range
attraction and a longer-range repulsion~\cite{16}. 
I have chosen physically and biologically relevant parameters as
justified above. I have observed a strong robustness of the cluster
phase with respect to the potential shape. I have tested repulsive
terms decaying linearly, algebraically (like $1/r^2$ or $1/r^4$
\cite{20}) or exponentially \cite{15} with the distance $r$ between
molecules, as well as attractive ones varying exponentially or
linearly with $r$. In all cases, a cluster phase exists at equilibrium
({\em i.e.} after very long runs) for wide ranges of
parameters. Clusters co-exist with a gas of monomers, of density
depending on potential and (weakly) on density $\phi$ (as observed
experimentally in Ref.~\cite{10}). What is important is not the
precise shape of the potential but the existence of a short-range
attraction of a few $k_BT$ and of a longer-range, weaker repulsion
extending on a range larger than the typical cluster
diameter. Therefore I have focused on a potential shape already
studied in detail \cite{15,Sear99,16}:
\begin{equation}
U(r) = - \varepsilon_a \exp(- \gamma_a r) + \varepsilon_r \exp(- \gamma_r
r). 
\end{equation}
The parameters are chosen so that, as required above, the binding
energy between two proteins is $-$4~$k_BT$ and the energy barrier is
0.1~$k_BT$. The following values fulfill this requirement:
$\varepsilon_a = 32$~$k_BT$ and $\varepsilon_r = 0.3$~$k_BT$;
$1/\gamma_a = 2$~nm and $1/\gamma_r = 16$~nm. In spite of the high
value of $\varepsilon_a$, the binding energy is low because the
attractive part is cut at $r=d_0$ due to the hard core repulsion
(Inset of Fig.~\ref{figure2}). This hard-core diameter is chosen as
$d_0 = 4$~nm, the typical diameter of a protein of average molecular
weight~\cite{10}.  The proteins are given a ``bare'' diffusion
coefficient $D_0 = 1$~$\mu$m$^2$/s \cite{18,note2}: at each Monte Carlo step
(MCStep), a randomly chosen protein attempts to move a distance
$\delta r$ forward in a randomly chosen direction; Here $\delta
r=1$~\AA $~\ll d_0$; With this $\delta r$, the acceptation rate of
MCSteps is larger than 60~\%, even at the highest densities
considered; The time step between two MCSteps is $\delta t = \delta
r^2 /(4D_0) = 2.5$~ns. A Monte Carlo sweep corresponds to $N$
MCSteps. The simulation time is chosen so that error bars on $D$ and
$\langle n \rangle$ are smaller than 10~\% (more than $10^7$ sweeps,
{\em i.e.}  30~ms of real time). The protein average long-term
diffusion coefficient $D$ is measured at different concentrations
$\phi = N d_0^2/a^2$ ($a$ is the size of the box with periodic
conditions in which the proteins diffuse). By long-term, it is meant
at time-scales larger than the time needed to diffuse inside the
clusters (typically 0.1~ms). The measures are performed after a long
equilibration period.  To be sure that equilibrium has indeed been
reached, I simulate two systems with initial configuration chosen as
(a) a random one where proteins later coalesce to form clusters; (b) a
completely condensed state where all proteins belong to the same big
cluster which later splits into smaller ones and gas. Equilibrium is
considered to have been reached when both systems (a) and (b) are
qualitatively identical (same number of multimeric clusters). The
Monte Carlo time needed is generally shorter than $10^7$ sweeps
(however, see~\cite{note}).  Note that the time needed to reach
equilibrium in (a) is rather long because after a transient period
where small clusters appear {\em via} a bimodal-like decomposition,
larger clusters are formed by evaporation of the smaller ones. 
Evaporation is the result of escape of single proteins from
the clusters, one after the other. The energy barrier to evaporate a single
protein being of several $k_BT$, this is a slow process~\cite{note}.

\section{RESULTS AND DISCUSSION}

\begin{figure}
\begin{center}
\ \psfig{figure=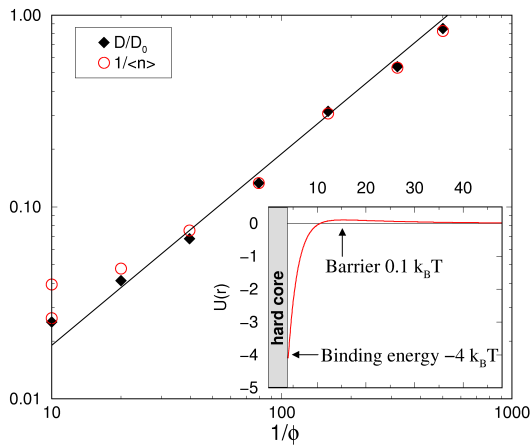,height=6cm} \
\end{center}
\caption{The diffusion coefficient $D/D_0$ (diamonds) and the inverse
mean cluster size $1/\langle n \rangle$ (circles) as a function of the
inverse density $1/\phi$, in log-log coordinates, for the potential
$U$ discussed in the text (see~\cite{note} for $\phi=0.1$). The full
line has slope 1, for comparison.  Inset: the potential $U(r)$ (in
units of $k_BT$; $r$ in nanometers).
\label{figure2}}
\end{figure}

As $\phi$ was increased, $D$ decreased dramatically, as expected
(Fig.~\ref{figure2}). One observes that $D \approx Const./\phi$ over
nearly two decades, thus confirming the hypothesis that cluster phases
can account for this behavior. This law had been observed 25 years
ago~\cite{18}, without receiving a full explanation, apart from
arguments invoking ``crowding effects'' or ``aggregation'' reminiscent
of the clustering mechanism discussed here, but unable to predict
quantitatively the dependence of $D$ on $\phi$~\cite{18}. The
diffusion coefficient at typical cell membrane protein concentration
($\phi \sim 0.1$) appears to be reduced by a factor larger than 10,
which is the decrease of $D$ observed in cell membranes as compared to
the same diffusion coefficient $D_0$ in model membranes at low $\phi$
\cite{6,18}.
One can also see in Fig.~\ref{figure2} that $\langle n \rangle$ is
proportional to $\phi$ at low concentration, as
expected. Interestingly, at higher $\phi$, the data for $\langle n
\rangle$ still follow this law for this choice of potential.  In
addition, the relation $D \simeq D_0/\langle n \rangle$ holds on all
the range of concentrations studied. This demonstrates that the
interactions between clusters are negligible and that they diffuse
independently at the time scale considered.

\begin{figure}[ht]
\begin{center}
\ \psfig{figure=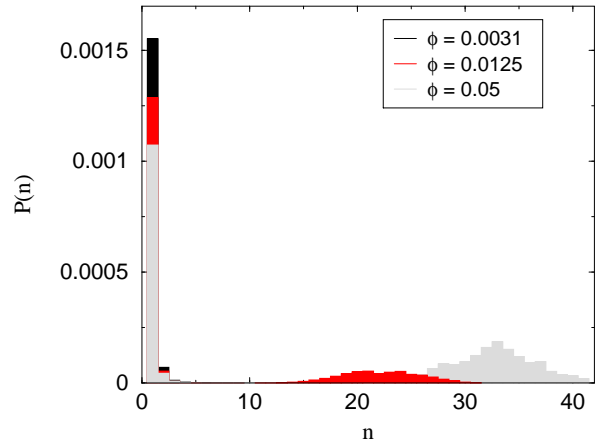,height=6cm} \
\end{center}
\caption{Numerical cluster-size distributions, for three different
concentrations: $\phi=0.0031$ (black), $\phi=0.0125$ (red) and
$\phi=0.05$ (gray). For $\phi=0.0031$, multimers are not visible with
the scale used because they are very scarce. The fractions $P(n)$ 
of clusters of size $n$ are
in the same units as $\phi$: number of clusters per unit
surface $d_0^2$.
\label{distrib}}
\end{figure}

As it was observed experimentally (Fig.~4 of Ref.~\cite{10}), the
cluster distributions obtained in simulations are bimodal on all the
range of concentrations where $\langle n \rangle \propto \phi$. A gas
of monomers ($n=1$) coexists with large multimers ($n>1$), the
distribution of which is Gaussian around a typical size $n^*$.  With
the parameters chosen here, there are virtually no small multimers
(dimers, trimers, etc\ldots).  These distributions are illustrated in
Fig.~\ref{distrib} at different concentrations $\phi$. The relation
$\langle n \rangle \propto \phi$ then comes from a subtle balance
between monomers and multimers: as $\phi$ increases, the density of
monomers is essentially constant while large clusters capture
additional proteins.

Once formed, clusters appear to be non-rigid, deformable,
fluctuating (as illustrated on the two consecutive snapshots in
Fig.~\ref{figure}, they can be seen as liquid droplets, that
reorganize rapidly; proteins diffuse inside the clusters with measured
short-term diffusion coefficients larger than $0.01~\mu$m$^2$/s) and
long-lived, as anticipated above. By long-lived, it is meant that
clusters are stable at the time-scale of the simulations (about
30~ms). However, some proteins constantly leave the clusters ({\em via}
the evaporation process discussed above), diffuse
freely in the gas phase and are captured later by another
cluster. Rare events of clusters being disintegrated or
nucleating spontaneously in the gas have been observed. Therefore
clusters are certainly not stable at long time scales (seconds or
minutes) and are only transitory.

I have also observed in simulations that small modulations of the
parameters $\varepsilon$ or $\gamma$ in $U(r)$, can lead to
segregation \cite{1,sieber06,19}: if two (or more) groups of proteins
(A's and B's, which are not necessarily identical in the same group)
are present in the simulation and if A-A and B-B interactions are
slightly favored as compared to A-B ones, then the proteins
segregate. There are A-rich and B-rich clusters because even though it
is entropically unfavorable, it is energetically favorable. For
instance, an A-B binding energy 10~\% smaller than equal A-A and B-B
ones suffices to ensure segregation. At the biological level, this
implies that groups of proteins which show a slight tendency to
associate, because of their physico-chemical properties, would
segregate in distinct clusters (see~\cite{sieber06}). This mechanism
might play an important role in sorting together proteins that must
congregate to perform biological functions~\cite{1,3}. This point will
be quantified in future investigations~\cite{Foret}.

The existence of cluster phases in live cells would mean that proteins
spontaneously congregate in small clusters of a few entities or a few
tens of entities in the plasma membrane instead of constituting a
``gas'' of independently diffusing inclusions, thus promoting
nano-domains in plasmic membranes. By contrast to the lipid raft
scenario where domains result from a lipidic micro-phase separation
and then recruit specific proteins, the present mechanism proposes
that domain formation is mainly driven by protein interactions (even
though lipids {\em do} play an important role in the effective
forces). Note that this scenario does not exclude a concomitant
recruitment by protein clusters of specific lipids having a higher
affinity for those proteins (and which participate in the effective
attractions~\cite{Gil98}), thus reconciling my hypothesis with the
observation of detergent-resistant membrane
fractions~\cite{Edidin03}. This mechanism, by constraining an
increasing fraction of lipids to diffuse slowly (with clusters) as
$\phi$ increases, could also explain why the diffusion constant of
{\em lipids} decreases significantly when the concentration of {\em
proteins} increases~\cite{18}.

In spite of the evidence provided above, the existence of protein
clusters has to be confirmed definitively at the experimental
level. Even though such clusters have already been observed by
freeze-fracture electron microscopy \cite{4,10}, their existence must
be explored by different techniques in a wider range of situations, in
cell and model membranes. Near-field scanning microscopy is an ideal
tool because it is able to identify individual proteins after
immobilizing the membrane onto an adequate substrate. Counting numbers
of proteins in clusters is then in principle
possible~\cite{Goncalves06}. Such experiments would be able to
investigate the dependency of cluster numbers $\langle n \rangle$ on
protein concentration, as well as the correlation between $\langle n
\rangle$ and $D$. Recently, high-frequency single-particle tracking
has demonstrated that proteins are confined in nano-domains in the
plasma membrane of live cells \cite{6}, of typical diameter a few tens
of nanometers. An appealing hypothesis is that these
nano-domains are the clusters under consideration. For proteins
several nanometers apart, the previous size would correspond to
clusters containing a few tens of proteins, in agreement with the
previous simulations. The confirmation of this hypothesis would
provide additional evidence of cluster phases in live cells. 

I have proposed in this paper a paradigm leading to the formation of
size-limited nano-domains in cell and model membranes. This scenario,
based on reasonable hypotheses about the energy and length scales in
biological membranes, sheds light on several so-far open issues in
cell biology: (i) it provides a mechanism for the limited size of
nano-domains; (ii) it gives a complete qualitative justification for
former experiments in model membranes~\cite{10,18}; (iii) it proposes
a simple explanation leading to the proportionality law $D \propto
\phi$. If cluster phases were to be experimentally confirmed in model
membranes and in live cells, it would mean that, by physical
mechanisms, proteins generically gather in small assemblies in
biological membranes, thus shedding new light on membrane functional
processes.

\medskip

\noindent{\bf Acknowledgments:} I am indebted to Lionel Foret and
Manoel Manghi for helpful discussions and comments.


\end{document}